\documentclass[aps,twocolumn,prb]{revtex4-1}

\usepackage{graphicx}
\usepackage{mathtools,amsthm,amssymb}

\usepackage[pdfauthor={Guido Burkard, V. O. Shkolnikov, and D. D. Awschalom},
pdftitle={A CPHASE gate between NV spins using cavity QED},
pdfsubject={},colorlinks=true,linkcolor=blue,citecolor=magenta]{hyperref}

\usepackage{bbm}

\begin{document}


\title{A cavity-mediated quantum CPHASE gate between NV spin qubits in diamond}

\author{Guido Burkard}
\author{V. O. Shkolnikov}
\address{Department of Physics, University of Konstanz, D-78457 Konstanz, Germany}

\author{D. D. Awschalom}
\address{Institute for Molecular Engineering, University of Chicago, Chicago, IL 60637, USA}



\begin{abstract}
While long spin coherence times and efficient single-qubit quantum control have
been implemented successfully in nitrogen-vacancy (NV) centers in
diamond, the controlled coupling of remote NV spin qubits remains challenging.
Here, we propose and analyze a controlled-phase (CPHASE) gate for the spins
of two NV centers embedded in a common optical cavity and driven by
two off-resonant lasers. 
In combination with previously demonstrated single-qubit gates, CPHASE allows 
for arbitrary quantum computations.
The coupling of the NV spin to the cavity mode is based upon
Raman transitions via the NV excited states and can be
controlled with the laser intensities and relative phase.
We find characteristic laser frequencies at which the scattering
amplitude of a laser photon into the cavity mode is strongly dependent
on the NV center spin. 
%
%
A scattered photon can be reabsorbed by another selectively
driven NV center and generate a conditional phase (CPHASE) gate.
Gate times around 200 ns are within reach, nearly three orders of magnitude 
shorter than typical NV spin coherence times of around 10 $\mu$s.     
The separation between the two interacting NV centers is only limited by the extension of the cavity.
\end{abstract}

\maketitle


\section{Introduction}
\begin{figure}[b]
\includegraphics[width=0.5\textwidth]{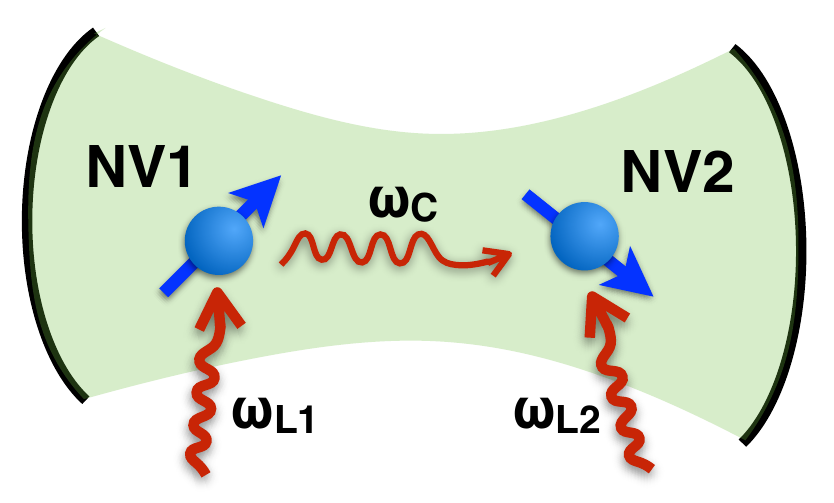}
\caption{Two nitrogen-vacancy centers in diamond located in an
optical cavity and coupled to a common cavity mode with frequency $\omega_C$  
(shown schematically).  
The NV centers are excited by off-resonant laser fields (frequencies $\omega_{Li}$).
Spin-dependent scattering of laser photons off the NV center into the
cavity mode and back allows for a coupling of the two NV spins which
produces the universal CPHASE quantum gate.}
\label{fig:system}
\end{figure}

Nitrogen-vacancy (NV) centers in diamond have emerged as powerful
and versatile quantum systems with applications as sources of
non-classical light, as high-precision sensors, and as qubits for quantum
information technology \cite{Dobrovitski2013}.
The electron spin of the NV center unites several essential properties required for quantum
information processing (QIP).  Its quantum coherence is preserved over
long times, even at elevated temperatures, and it allows for optical
preparation and read-out, as well as quantum gate
operations via radio-frequency (rf) excitation, at the level of a
single-NV center.
One of the remaining challenges on the way towards diamond-based 
QIP is the establishment of a scalable architecture allowing for the
coherent coupling between NV spins.   
A controlled coupling is required to realize a two-qubit gate such as
controlled-phase (CPHASE) or controlled-not (CNOT) which
forms a universal set of quantum gates in combination with
single-qubit gates.
Controlled operations between the NV electron spin and a nearby
nuclear spin have been performed using a combination of rf and 
microwave pulses \cite{Gaebel2006}, whereas
entanglement generation can be achieved between the electron spins of 
two nearby NV centers on the basis of static dipolar interactions \cite{Dolde2013},
and between NV center spins separated by several meters
\cite{Pfaff2012}, and subsequently over more than one kilometer
\cite{Hensen2015} via a non-deterministic coincidence measurement protocol.
Here, we propose and theoretically analyze a fully controllable and
switchable coupling between the spins
of distant NV centers coupled to the same mode of a surrounding
optical cavity (Fig.~\ref{fig:system}).

A variety of optical cavity systems for cavity quantum electrodynamics
(QED) coupled to   defect centers in diamond exist.
The advantage of whispering gallery modes of silica microsphere is
their ultrahigh quality factors \cite{Park2006} $Q > 10^8$,
whereas photonic crystals fabricated within the diamond crystal 
\cite{Wang2007,Faraon2012,Riedrich2012} or on top 
\cite{Englund2010} allow for the embedding of the NV centers directly
into the optical cavity structure, but comprise (so far) somewhat
lower $Q$ factors.  However, photonic crystal cavities in diamond
with $Q > 10^5$ have recently been fabricated \cite{Burek2014}.
The architecture to be proposed here can in principle be used with any
realization of NV-cavity coupling, provided sufficiently high $Q$
and dipole matrix element of the 
ground state (GS)-excited state (ES) transition in the cavity field.

\begin{figure}
\includegraphics[width=0.5\textwidth]{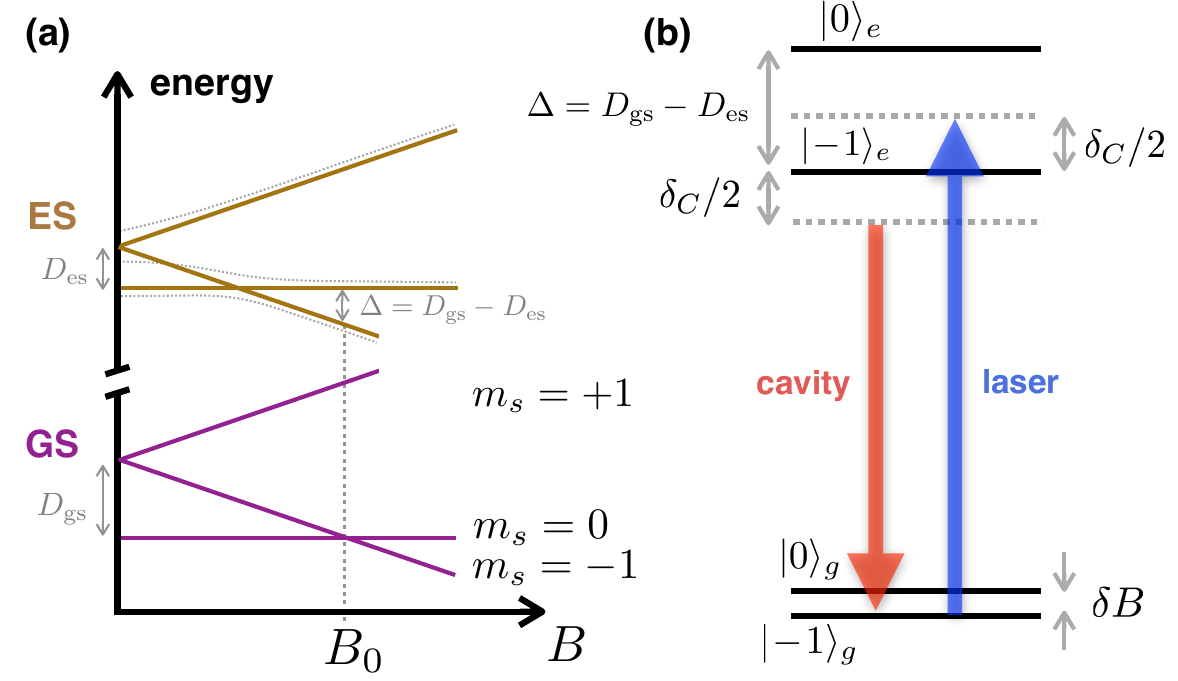}
\caption{(a) GS and ES energy levels as a function of the 
magnetic field $B$ applied along the NV axis. 
In the ES, only one orbital triplet is shown.
The effect of the spin-spin interactions $\Delta_{1,2}$ is shown
schematically by the dotted lines.
(b) Simplified energy level scheme.  Here,
$|0\rangle_g$ and $|-1\rangle_g$ denote orbital ground-state levels 
with spin projection $m_s=0$ and $m_s=-1$. Similarly,
$|0\rangle_e$ and $|-1\rangle_e$ stand for the corresponding 
excited-state levels.  The scattering of a laser photon (blue) into a
cavity photon (red) via the intermediate excitation of the NV center is
suppressed in the $m_s=-1$ state by destructive quantum interference 
when $\delta_L=\Delta +\delta_C/2$.}
\label{fig:level_scheme}
\end{figure}

The basic working principle of the quantum gate operation proposed
here is as follows.  We restrict ourselves to two of the three 
GS spin triplet states, $m_s=0$ and $m_s=-1$, which will serve as the qubit
basis in our scheme  (Fig.~\ref{fig:level_scheme}a).  
Near the GS level crossing around a magnetic
field of about $B_0\sim 1000\,{\rm G}$, these two states are nearly
degenerate, and separated by several GHz from the third ($m_s=+1$)
state.  Off-resonant coupling of the GS-ES transition to the cavity
mode combined with off-resonant laser excitation can be used to
generate Raman-type two-photon transitions starting and ending
in the GS, accompanied by the scattering of a laser photon into the 
cavity mode, or vice versa (Fig.~\ref{fig:level_scheme}b).
The off-resonant coupling is the main distinguishing feature from
resonant schemes which are limited by spontaneous emission
\cite{Zagoskin2007}.
The proposed two-qubit coupling mechanism relies on a 
spin-dependent scattering of laser photons into the cavity and back
which is possible because of the difference in zero-field splittings
in the GS and ES. 
More specifically, the $m_s=0$ and $m_s=-1$ states in the ES are 
not degenerate at $B_0$, which leads to unequal scattering matrix 
elements for the $m_s=0$ and $m_s=-1$ states.  To produce an
entangling quantum gate between two NV spin qubits, we find it to
be sufficient if the laser-cavity photon scattering rate is different
for the two spin states.   
If two NV centers are simultanously
coupling in this way to the same cavity mode, they will exchange
a virtual cavity photon, thus generating a conditional phase shift;
once the accumulated relative phase amounts to $\pi$, 
a CPHASE gate on the two NV spin qubits has been achieved.

In contrast to cavity-mediated spin interactions proposed 
for semiconductor quantum dots \cite{Imamoglu1999} 
where the spin-orbit splitting in the valence band can be
used for spin-selective excitation with polarized radiation
and Raman-type spin flip transitions, we propose here
to use another mechanism based on the different zero-field splittings of the
NV ground and excited states to perform phase and 
controlled-phase operations.
Earlier work on cavity-mediated quantum gates for 
defect qubits in diamond makes
use of spectral hole burning \cite{Shahriar2002} or 
a series of $\Lambda$ systems\cite{Solenov2013}. The latter requires
a sequence of at least three two-color pulses, while
our scheme manages on just one single-color laser pulse
for a CPHASE gate.
A model for three NV centers coupled to a whispering-gallery 
mode in a silica microsphere cavity using polarized excitation
has been studied with the goal of achieving a three-qubit
CPHASE gate \cite{Yang2010}.  Our scheme relies on spectral selectivity
and thus does not require polarized excitation.  The effect studied
here produces an elementary, universal two-qubit CPHASE gate.
%


\section{Single NV center in a cavity}
The NV center in its ground state (GS) and excited state (ES) spin
triplet will be described by the Hamiltonian
\begin{equation}
H_{\rm NV} = g_e \mu_B B S_z 
+\left(\begin{array}{c c}
  E_g + D_{\rm es}  S_z^2     &  g_L^* e^{-it\omega_L} \\
g_L e^{it\omega_L}  & D_{\rm gs} S_z^2 
\end{array}\right),
\label{eq:HNV}
\end{equation}
%
%
where the first term describes the Zeeman splitting of the 
spin ${\bf S} = (S_x,S_y,S_z)$ with eigenvalues $m_s=-1,0,1$ in a
magnetic field applied along the NV ($z$) axis with identical electronic
Land\'e g-factor $g_e$ for the GS and ES ($\mu_B$ denotes the Bohr magneton).
See Appendix \ref{sec:alignment} for a discussion of a possible magnetic field
misalignment.
The second term in Eq.~(\ref{eq:HNV}) includes the GS-ES energy gap
$E_g=1.945\,{\rm eV}$
and the distinct GS and ES zero-field spin splittings $D_{\rm gs}=2.88\,{\rm GHz}$
and $D_{\rm es}=1.44\,{\rm GHz}$.  
The off-diagonal terms describe laser excitation at a frequency
$\omega_L$, with the spin-independent dipole matrix element $g_L$.  
We assume that the ES orbital
state energies are strongly split by the strain in the diamond
crystal, and we can concentrate on one of the two orbital ES
triplets.  The prerequisite for this to be a reasonable approximation 
is that the strain splitting exceeds the ES spin-orbit coupling
$\lambda=5.3\,{\rm GHz}$.
Strain splittings in excess of this value and up to 20 GHz 
have been observed \cite{Batalov2009,Bassett2014}.
Taking only one orbital ES into account, we can view the Hamiltonian
$H_{\rm NV}$ in
Eq.~(\ref{eq:HNV}) as a 6x6 matrix consisting of four 3x3 blocks. 
The Zeeman splitting described by the first term in Eq.~(\ref{eq:HNV})
is independent of the orbital state.
Using Pauli matrices $\tau_i$ to describe the GS-ES orbital state, 
i.e., $\tau_z=-1$ for the GS and $\tau_z=+1$ for the ES,
and working in a rotating frame with the frequency $\omega_L$, 
we can write 
\begin{equation}
H_{\rm NV} =  g\mu_B B S_z + D S_z^2 - \frac{\Delta}{2}  S_z^2 \tau_z
+ \frac{\delta_L}{2}\tau_z + g_L\tau_- + g_L^*\tau_+,
\end{equation}
where $D=(D_{\rm gs}+D_{\rm es})/2 =2.16\,{\rm GHz}$ and $\Delta=D_{\rm gs}-D_{\rm
  es}=1.44\,{\rm GHz}$
denote the mean and difference between the GS and ES
zero-field splittings, $\tau_\pm=(\tau_x\pm i \tau_y)/2$ describe
transitions between the GS and ES,  
and $\delta_L = E_g - \omega_L$ is the laser detuning.
We have so far neglected the spin-spin couplings in the ES,
but will discuss their effect further below.

We now consider a single NV center coupled to a near-resonant
mode of a surrounding optical cavity which we describe, using the
rotating-wave approximation, with the following Hamiltonian,
\begin{equation}
H =  H_{\rm NV} + \delta_C a^\dagger a 
+ g_C \left( \tau_+ a +  \tau_-  a^\dagger \right),
\label{H}
\end{equation}
where $\delta_C = \omega_C-\omega_L$ denotes the detuning  of the cavity mode
from the laser excitation frequency
and $a^\dagger$ ($a$) creates (annihilates) a cavity photon.
The dipole matrix element $g_C$ of the cavity field can be
made real-valued by an appropriate phase convention in the 
excited state.  However, $g_L$ can in general not be made 
real-valued at the same time; its phase $\phi$ depends on the phase
of the laser field.

The magnetic field is chosen at a working point around the 
GS level crossing $B_0=D_{\rm gs}/g_e\mu_B$ where we focus our
description on the nearly degenerate $m_s=-1$ and $m_s=0$ levels (the $m_s=+1$
level will be included further below).
This approximation is justified because the $m_S=+1$ level is 
split off by the zero-field splitting which is much larger than the 
spin-spin splittings coupling it to the other two spin levels.
We describe here the situation of an initially empty cavity, which 
subsequently holds at most one virtual photon.
Starting from an empty cavity, and assuming 
sufficiently large detunings $\delta_C \gtrsim g_C$ 
and $\delta_L \gtrsim g_L$ of the cavity and laser frequencies, we can further reduce
the relevant states to 
$|G0\rangle=|G,n=0\rangle$, $|G1\rangle=|G,n=1\rangle$, 
and $|E0\rangle=|E,n=0\rangle$, where
$G$ and $E$ denote the GS and ES, respectively, and $n$ denotes
the cavity photon number. Including the two remaining spin
projections, $m_s=-1,0$ this leaves us with six states for a single NV
and the cavity.

The combined action of the coupling to the laser and cavity fields
can scatter a photon from the laser into the cavity or vice versa,
via an intermediate virtual ES.  Starting from the Hamiltonian Eq.~(\ref{H}),
and assuming that the electric dipole couplings $g_{L,C}$ are much smaller than
the detuning from the one-photon resonances, we can derive an effective GS Hamiltonian for such second-order
processes (see below),
\begin{equation}
\tilde{H} = \delta_C a^\dagger a  
+\delta B |0\rangle\langle 0|
+ \sum_{m_s=0,-1} \!\!\!  g_{m_s}|m_s\rangle\langle m_s| \left(g a+g^* a^\dagger\right),
\label{eq:effH1}
\end{equation}
where $ |m_s\rangle\langle m_s|$ denotes the projection operator on the 
spin state with projection $m_s$, 
\begin{equation}
g_{m_s} =  - g_L g_C \frac{\delta_L - \delta_C/2 + m_s\Delta}
{(\delta_L  + m_s\Delta )( \delta_L-\delta_C +m_s\Delta )}
\label{eq:g}
\end{equation}
the effective coupling strength, and $\delta B = B - D_{\mathrm{gs}}/g_e\mu_B$
the magnetic field detuning from the GS level crossing.
The last term in Eq.~(\ref{eq:effH1}) describes spin dependent 
scattering processes at the NV center
of  a cavity  photon into a laser photon or vice versa.
Generally, we find that in order to construct a CPHASE gate,
it is sufficient if $g_{0} \neq g_{-1}$ (see also below).
A possible extreme case where $g_{0} =0$ is described 
in the Appendix \ref{sec:toy}.
In Eq.~(\ref{eq:effH1}), we have suppressed optical Stark and
Lamb shifts of order $g_L^2$ and $g_C^2$, which will not play an essential role in
what follows.

We now give a more detailed derivation of Eqs.~(\ref{eq:effH1}) and (\ref{eq:g}),
starting from Eq.~(\ref{H}).
To describe the combined action of the coupling between the NV center to the laser and cavity fields
we write Eq.~(\ref{H}) as $H=H_0+V$ with the perturbation Hamiltonian
\begin{equation}
V=g_L\tau_- + g_L^*\tau_+ + g_C (\tau_+ a+\tau_-a^\dagger),
\end{equation} 
and eliminate the ES in order to derive an effective interaction using the 
Schrieffer-Wolff transformation\cite{DVincenzo,Winkler}, 
\begin{equation}
H_{\rm eff}=e^{S} H e^{-S} = H_0 + \frac{1}{2}[S,V]+\cdots,
\label{Single_NV_in_a_cavity_S_matrice}
\end{equation}
generated by the antihermitian operator
\begin{eqnarray}
S &=& -g_L \left(\delta_L - \Delta +\Delta |0\rangle\langle
  0|   \right)^{-1} |G0\rangle\langle E0|\\
 & & -g_C \left(\delta_L -\delta_C- \Delta  +\Delta |0\rangle\langle
  0|\right)^{-1}|G1\rangle\langle E0| 
-\mathrm{h.c.},\nonumber
\end{eqnarray}
such that
$[S,H_0] = -V$,
and obtain the effective GS interaction Hamiltonian
\begin{eqnarray}
\tilde{H} &=& H_0+\frac{1}{2} [S,V]\Big|_{\rm GS} ,
\label{eq:effH1M}
\end{eqnarray}
which directly leads to Eqs.~(\ref{eq:effH1}) and (\ref{eq:g}).


\section{Two NV centers coupled to a common cavity mode}
\label{sec:two NVs in a cavity}
The scattering of a photon from the laser to the cavity
field and vice versa, conditional on the spin (qubit) state of an NV
center can be used to construct a cavity-photon mediated quantum
gate between two NV spin qubits coupled to a common cavity mode.
Starting from two NV centers ($i=1,2$), each coupled to the same cavity mode as 
described above (Fig.~\ref{fig:system}), we derive the effective
coupling Hamiltonian for two NV spins by eliminating the virtual
cavity photon.

It is important to recognize that the cavity mediated interaction between
the NV centers is a fourth-order process in the coupling strengths which prevents us from using
the second-order Hamiltonian Eq.~(\ref{eq:effH1}) directly to 
calculate the coupling between the NV center spins.
In order to systematically account for all contributions up to the
fourth order, we perform a fourth-order Schrieffer Wolff
transformation of the Hamiltonian describing two NV centers coupled to
a common cavity mode, 
\begin{eqnarray}
H &=&H_0+H_{\rm int},\nonumber \\
H_0 &=&\delta_Ca^\dagger a +
        \sum\limits_{i=1,2} \left[ \frac{1 + \tau_{z}^i}{2}
\left(\delta_{Li}+\Delta S_{zi} \right) + \delta B_i S_{zi} \right],\nonumber \\
H_{\rm int} &=& \sum\limits_{i=1,2} \left( g_{Li}\tau_{-}^i +
                g_{Ci}a^\dagger\tau_{-}^i+h.c. \right) ,
\label{Two_NVs_Hamiltonian}
\end{eqnarray}
where we have restricted ourselves to the $S_z=0$ and $S_z=-1$ states
near the GS level crossing where $S_z^2 = -S_z$.
As this Hamiltonian commutes with the operators $S_{z1}$ and $S_{z2}$ of the NV centers, we can
treat it separately for each of the four ground-state spin
configurations,  which represent the logical basis for our two-qubit system.
For each spin configuration, we consider the five states 
$|GG0\rangle$, $|GG1\rangle$, $|EG0\rangle$, $|GE0\rangle$, and $|EE0\rangle$,
where $|X_1 X_2 n\rangle$ denotes the state with NV $i$ ($i=1,2$) in the ground
($X_i=G$) or excited ($X_i=E$) state, while the cavity mode is occupied
with $n$ photons.
In analogy with the previous section we are only interested in the effective
interaction between the NV centers and the cavity in the NV ground state.
To derive an effective spin Hamiltonian for the NV ground states, we
decouple the two states $|GG0\rangle$ and $|GG1\rangle$ from the
remaining three states by performing a Schrieffer-Wolff transformation\cite{DVincenzo,Winkler}.
In analogy with Eq.~(\ref{Single_NV_in_a_cavity_S_matrice}), and
expanding to fourth order, we have
\begin{eqnarray}
H_{\rm eff}&=&e^{S} H e^{-S} = H +[S,H]+ \frac{1}{2}[S,[S,H]]+ \nonumber \\
&+&\frac{1}{6}[S,[S,[S,H]]]+\frac{1}{24}[S,[S,[S,[S,H]]]]. 
\label{Two_NVs_S_matrice}
\end{eqnarray}
We then expand the matrix S as a series $S=S_1+S_2+S_3+S_4+ \ldots$, 
where each term $S_i$ is derived using Eq.~(\ref{Two_NVs_S_matrice}) 
under the requirement that there is no coupling between the
$|GGn\rangle$ ($n=0,1$)
subspace and the excited states of the NV centers up to $i$-th order
in the coupling constants $g_L$ and $g_C$.
In the sum Eq.~(\ref{Two_NVs_S_matrice}), we then calculate all the
residual terms and obtain the effective Hamiltonian in the basis $|GG0\rangle$, $|GG1\rangle $,
\begin{equation}
H_{\rm eff} =\left(\begin{array}{c c}
  W_{GG0}+|\tilde{g}|^2/\delta_C     &  \tilde{g}^* \\
\tilde{g} & W_{GG1}-|\tilde{g}|^2/\delta_C 
\end{array}\right).
\label{Two_NVs_Effective_Hamiltonian}
\end{equation}
Introducing the phases $\phi_i$ of the lasers as $g_{Li}=|g_{Li}|e^{i\phi_{i}}$, 
we find for the eigenenergies of this effective Hamiltonian 
\begin{eqnarray}
W_{GG0} &=&\sum_{i=1,2} \left[
\delta B_im_{si} 
- \frac{|g_{Li}|^2}{\delta_{Li}+\Delta m_{si}}
\right. \nonumber \\
& & \left. +\frac{|g_{Li}|^4}{(\delta_{Li}+\Delta m_{si})^3} - \frac{|g_{Li}|^2|g_{Ci}|^2}{(\delta_{Li}+\Delta
    m_{si})^2\delta_C} \right] \nonumber \\
& & - \frac{2|g_{L1}g_{L2}|g_{C1}g_{C2}\cos{(\phi_1-\phi_2)}}{(\delta_{L1}+\Delta m_{s1})(\delta_{L2}+\Delta m_{s2})\delta_C} ,
\label{Two_NVs_Ground_state_energy}
\end{eqnarray}
and $W_{GG1}   \approx \delta_c+\sum_{i=1,2}\delta B_im_{si} $, whereas for the off-diagonal matrix
element we obtain
\begin{equation}
\tilde{g}  =
-\sum_{i=1,2}\frac{e^{i\phi_i}g_{Ci}|g_{Li}|(\delta_{Li}+\Delta  m_{si}-\delta_C/2)}
{(\delta_{Li}+\Delta m_{si})(\delta_{Li}+\Delta m_{si}-\delta_C)}.
\end{equation}
We present $W_{GG0}$ only up to the fourth order corrections, as only
these terms will be important for the following discussion. 
We have also calculated $W_{GG0}$
using conventional perturbation theory, rather than a Schrieffer-Wolff
transformation, with identical results (see Appendix
\ref{sec:perturbation-analysis}).
The expression for $W_{GG0}$ in
Eq.~(\ref{Two_NVs_Ground_state_energy}) 
consists of two parts, where each term of the first part depends 
on the spin state of only one NV center and thus
only leads to single qubit dynamics.
Entanglement can be generated by the second part (last term) of
Eq.~(\ref{Two_NVs_Ground_state_energy}),
\begin{equation}
\epsilon_{m_{s1},m_{s2}}=-\frac{2|g_{L1}g_{L2}|g_{C1}g_{C2}\cos{(\phi_1-\phi_2)}}{(\delta_{L1}+\Delta m_{s1})(\delta_{L2}+\Delta m_{s2})\delta_C} 
\label{Two_NVs_Ground_state_entangling energies},
\end{equation}
as it depends on the spin state of both NV centers. 
Calculating this term for each spin configuration
$|m_{s1},m_{s2}\rangle= |-1,-1\rangle, |-1,0\rangle, |0,-1\rangle,|0,0\rangle $
leaves us with the diagonal spin Hamiltonian
\begin{equation}
H_{\rm 2q} =\left(\begin{array}{c c c c}
\epsilon_{-1,-1}     &  0  &  0 &  0 \\
               0                   & \epsilon_{-1,0} & 0 & 0 \\
               0                   &  0  & \epsilon_{0,-1}  & 0 \\
               0                   &  0  &  0  &  \epsilon_{0,0}
\end{array}\right).
\label{Two_NVs_spin_hamiltonian}
\end{equation}
This Hamiltonian generates a quantum gate $\exp(-itH_{\rm 2q}) $ which up 
to single qubit operations, is the CPHASE gate $U={\rm diag}(1,1,1,e^{i\gamma})$ with
\begin{equation}
\gamma = \frac{2|g_{L1}g_{L2}|g_{C1}g_{C2}\Delta^2\cos(\phi_1-\phi_2)}
{\delta_C\delta_{L1}\delta_{L2}(\delta_{L1}-\Delta)(\delta_{L2}-\Delta)}t .
\label{Two_NVs_CPHASE}
\end{equation}

Equation~(\ref{Two_NVs_CPHASE}) proves that the interaction of
two NVs through the cavity can give rise to an entangling gate. This
gate can be controlled both by the amplitude $|g_{Li}|$ and phase
$\phi_i$ of the lasers and by the detuning of the laser frequency from
the cavity mode $\delta_{Li}$.

The results of this section can only be considered a qualitative proof
of the entangling gate. They are valid as long as the perturbation
analysis works, which implies that the couplings $g_{Li},g_{Ci}$ are
much smaller than the detunings $\delta_C,\delta_{Li}$. Moreover, to
make predictions one should take into account the spin-spin
interaction in the excited state of the NVs, which will be done in the
next section in the description of our numerical results.


\section{Spin-spin interaction}
To make quantitative predictions, we need to include the spin-spin
interactions in the ES which have been studied both experimentally \cite{Fuchs2008,Batalov2009,Bassett2014}
and theoretically \cite{Doherty2011,Maze2011},
\begin{equation}
H_{s}=\frac{1}{2} \left(1 + \tau_z \right)\left[
\frac{\Delta_1}{2}\left(S_y^2-S_x^2\right)
+\frac{\Delta_2}{\sqrt{2}}  \left(S_x S_z+S_z S_x\right)
\right],
\end{equation}
where $\Delta_1=1.55\,\mathrm{GHz}$ and $\Delta_2\simeq 0.15\,\mathrm{GHz}$.
%

The Hamiltonian of the system will then take the form 
\begin{equation}
H=H_0+H_{\mathrm{int}}+H_{s},
\end{equation}
where $H_0$ and $H_{\mathrm{int}}$ have been introduced in the previous section. In the spin Hamiltonian the $\Delta_1$ term mixes the spin states $\mathrm{m_s}=-1$ and $\mathrm{m_s}=1$, while the $\Delta_2$ term
mixes $\mathrm{m_s}=-1$ and $\mathrm{m_s}=0$, as well as $\mathrm{m_s}=0$ and $\mathrm{m_s}=1$. Therefore, we can no longer treat each of the four logical states separately.

It is important to note that both cavity photon creation 
and spin-spin interaction are only possible when one of the NVs is in the excited state. To achieve this and thus create a quantum gate, laser excitation can be used to transform the initial ground state of the NVs. But it is also important that after the excitation is switched off, the system should remain in a final state that is the coherent 
superposition of the logical basis states. Thus the probability to have an excited NV after the laser pulse is turned off should be very low. This will be the case if the intensity of the lasers changes slowly, such that the adiabatic theorem provides that the system remains in the same eigenstate of the time-dependent Hamiltonian.
The final state of the system after the pulse is turned off will correspond to the ground state of the NVs and zero cavity photons $\textendash$ the logical basis of the two qubit system.

\begin{figure}[hbt]
\includegraphics[width=0.5\textwidth]{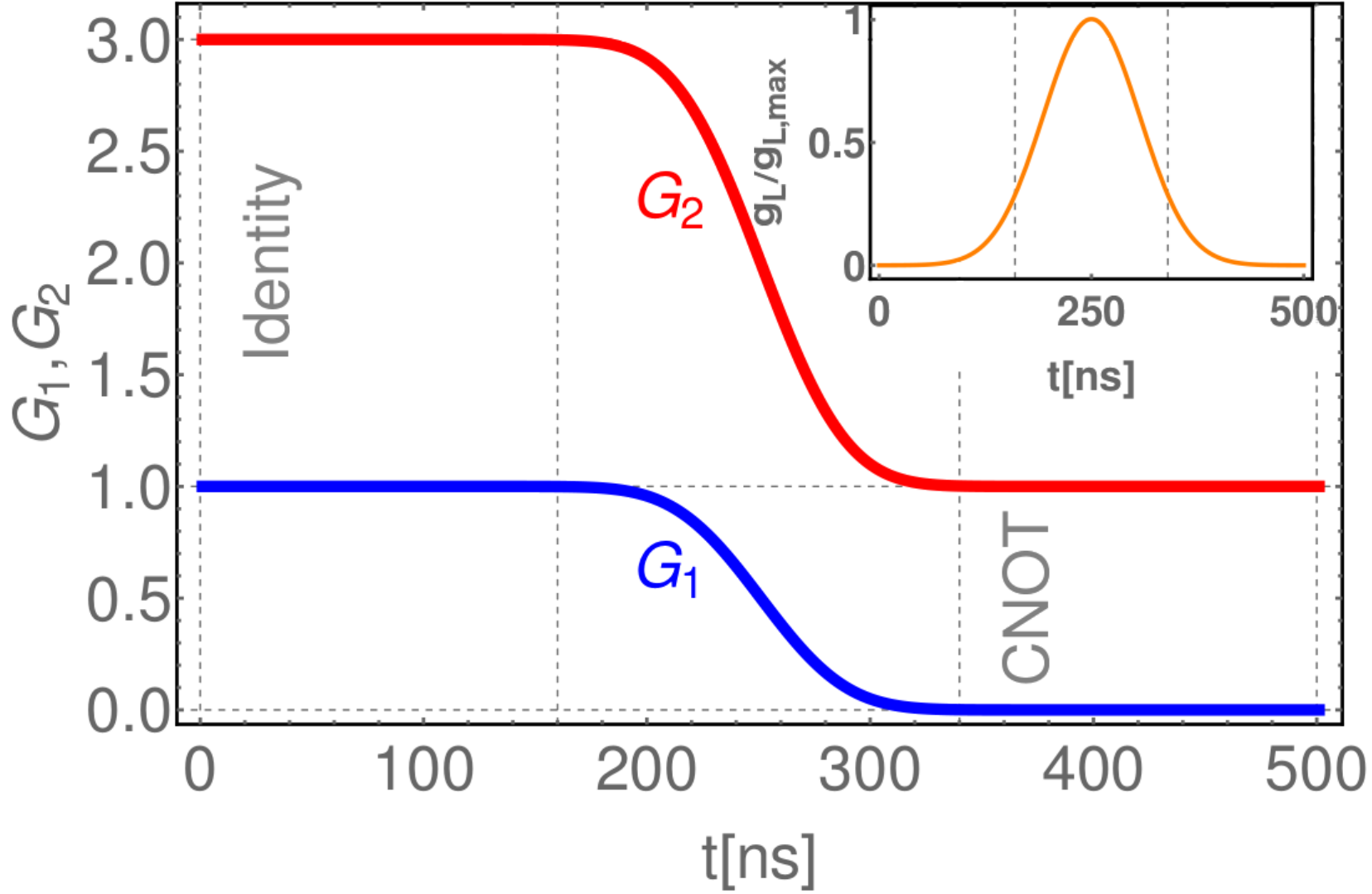}
\caption{Time dependence of the Makhlin invariants $G_1$ and $G_2$
  during the operation of a cavity mediated two qubit gate. 
Before the lasers are turned on $G_1=1\,\,\mathrm{and}\,\,G_2=3$, which corresponds to the identity operation. 
When the lasers are turned on, the two NVs start to interact through
the cavity, which leads to the appearance of entanglement and change
of Makhlin invariants. After the lasers are turned off the final state
of the system is related to the initial one by a CNOT operation, 
characterized with Makhlin invariants
$G_1=0\,\,\mathrm{and}\,\,G_2=1$.  
The parameters chosen for this plot are 
$g_C=100\,{\rm MHz}$, $\delta_C=400\,{\rm MHz}$, 
$\delta_L=1640\,{\rm MHz}$.
Inset: Laser pulse shape with maximum $g_{L,{\rm max}}=24\,\mathrm{MHz}$.}
\label{spin_spin_Makhlin_invariants}
\end{figure}

We now introduce
our numerical results obtained for this system, including spin-spin interactions. The laser detuning $\delta_L$ and the cavity detuning $\delta_C$ are asumed to be $1640\,\mathrm{MHz}$ and $400\,\mathrm{MHz}$ respectively.
The distance between the ground state 
and the lower excited state of the NV would then be $\delta_L-\Delta=200\,\mathrm{MHz}$ for $m_s=-1$ spin state and $\delta_L=1640\,\mathrm{MHz}$ for $m_s=0$ spin state. The energy of the cavity excitation would be $\delta_C=400\,\mathrm{MHz}$. 
The inverses of these values ($5\,\mathrm{ns},\,0.6\,\mathrm{ns},\,2.5\,\mathrm{ns}$ respectively) define the internal dynamical rate of the system, with respect to which one has to choose the ramp time of the pulse.
To stay within the adiabatic regime we took the pulse $g_L(t)$ to be a
convolution of a Gaussian and a rectangle with the widths
$133\,\mathrm{ns}$ (FWHM) and  $20\,\mathrm{ns}$ respectively. The
coupling $g_L$ at the maximum of the pulse is assumed to be 
$g_{L,{\rm max}}=24\,\mathrm{MHz}$. The coupling between the NV and the cavity is assumed to be $g_C=100\,\mathrm{MHz}$. 
We consider both NVs to be identical and driven by two
identical and synchronized lasers with the same amplitude, phase and
the pulse form described above.  Note that the two-qubit gate
operation requires neither the NV centers nor the driving fields to be
identical; this choise is made here only to simplify the analysis.
Under these assumptions we numerically
propagate each of the four logical states of the system. 
This results in a $4\times4$ unitary in the logical space of the two-qubit system, corresponding to
a CNOT gate, as shown by the Makhlin invariants $G_1$ and $G_2$ (Fig. \ref{spin_spin_Makhlin_invariants}),
for which the values 1 and 0 respectively were obtained, which is a characteristic of a CNOT gate\cite{Makhlin2002}.


\section{Discussion}

We have shown that virtual exchange of photons in an optical cavity
can mediate the two-qubit CPHASE gate between two NV spin qubits
in diamond.  Combined with
single-qubit operations, produced by rf excitation or by
laser fields \cite{Yale2013}, the CPHASE
gate allows for arbitrary (universal) quantum computations.
Therefore, optical cavity QED with NV centers in diamond represents
a realistic path towards spin-based quantum information processing.
The cavity-mediated quantum gate proposed here could be applied 
to other defects with a similar level structure, i.e., comprising 
spin triplet ground and excited states with deviating zero-field 
splittings.  For example, we expect that the gate protocol would 
also work for certain divacancy centers in silicon carbide.

As a further prerequisite for the scheme to work, the NV spin
coherence time and average time between cavity photon loss
must be longer than the gate operation time $t \sim 200\,{\rm ns}$.
The NV spin coherence time can reach
$1/\gamma_2 = T_2 \sim 10\, \mu s$, even at elevated temperatures.  
The photon loss rate can be estimated as 
$\tau^{-1} \sim (g/\delta_C)^2   \omega_C/ 2\pi Q$ 
where $(g/\delta_C)^2\sim 10^{-3}$ is the probability for the cavity mode
to be occupied by a virtual photon during the gate operation, and 
$\kappa = \omega_C/ 2\pi Q$ is the photon loss rate in the cavity with quality
factor $Q$.  For the parameters used above, a Q factor of $Q\sim 10^5$
is needed to achieve $\tau\sim 200\,{\rm ns}$. 
Because $\tau \sim \delta_C^2$ while $t \sim 1/g_{12}\sim \delta_C$,
increasing the detuning $\delta_C$
allows the use of cavities with lower $Q$ at the expense of slower gates, which
in turn are admissible for sufficiently long $T_2$.  The limit of
this scaling can be described in terms of a (coherent) cooperativity
factor \cite{Kolkowitz2012} $C_2 = g/\sqrt{\kappa\gamma_2} \gg 1$.

Finally, we expect this scheme to work below a temperature of 
about $20\,{\rm K}$ where the excited state levels are stable.
It is an open question whether a variation of this scheme will
also work at higher temperatures.

In a scalable qubit architechture, pairs of qubits need to be
selectively coupled within a large array.  A possible architecture comprises
single NV centers in optical cavities linked via optical fibers
\cite{Nemoto2014}.  The coupling mechanism described here lends
itself to another architechture where many NV centers are embedded in
a single cavity. In an array with
separations between NV centers on the order of 10 to 100 nm, 
selective pairwise coupling can be accomplished with a combination of spatial and spectral 
selectivity of the laser excitation.


\section*{Acknowledgments}
We thank Adrian Auer, Christopher Chamberland, Mikhail Lukin,
and Chris Yale for helpful discussions. 
We acknowledge funding from
AFOSR and NSF (DDA), 
and from CAP, DFG SFB767, and BMBF Q.com-HL (GB).


\appendix

\section{Magnetic field alignment}
\label{sec:alignment}

In our model, we have so far assumed that the magnetic field is
perfectly aligned with the NV axis of both defects involved in the
CPHASE gate.  This raises two important issues: (1) how to treat 
NV centers with different oriantations with respect to the diamond
crystal, and (2) to what extent will the CPHASE operation be disturbed
by any small misalignment of the magnetic field?  As for (1), we note
that there are four distinct NV orientations (up to small
misalignments which we discuss below).  Only the NV centers with
their orientation along the external B field will be near resonance
and will participate in the CPHASE gate operation while the NV centers
oriented along the three other axes can be safely ignored.  Regarding
(2), the field misalignment will add a term $g\mu_B B_x S_x$ to the
Hamiltonian Eq.~(\ref{eq:HNV}) where $B_x =  B\tan \phi \approx B
\phi$ is the transverse (misalignment) field (chosen to point in
$x$ direction) and $\phi \ll 1$ denotes the misalignment angle.  
The effect of the misalignment field is small if $B_x \ll \delta B$.
For a misalignment of one degree, the NV center should be operated 
at least $\delta B \approx 20\,\textrm{G}$ away from the level anticrossing.

\section{Minimal model for spin-dependent cancellation of laser-cavity
  photon scattering}
\label{sec:toy}

Here, we provide a minimal model to explain the spin-dependent cancellation of laser-cavity
  photon scattering.
Neglecting spin-spin coupling and assuming $g_L$ and $g_C$ to be real, 
we can treat the two spin states
$m_S=0$ and $m_S=-1$ separately, with the Hamiltonian 
\begin{equation}
H(m_S) = \left(\begin{array}{c c c}
0 & 0 & g_L\\
0 & \delta_C & g_C \\
g_L & g_C & \delta_L +m_S\Delta
\end{array}\right),\label{eq:bareH}
\end{equation}
in the basis $|G0\rangle$, $|G1\rangle$, $|E0\rangle$.
For $\delta_L = \Delta +\delta_C/2$, we find for 
the $m_S=-1$ state,
\begin{equation}
\label{eq:Htoy}
 H({m_S=-1}) = \left(\begin{array}{c c c}
0 & 0 & g_L\\
0 & \delta_C & g_C \\
g_L & g_C & \delta_C/2
\end{array}\right).
\end{equation}
Note that in the rotating frame, the excited state now lies 
exactly in between the states with zero and one cavity photon.
We introduce the dressed states $|\tilde{X}\rangle = e^S |X\rangle$,
\begin{eqnarray}
|\tilde{G0}\rangle  &=&  \left(1-\frac{2
    g_L^2}{\delta_C^2}\right)|G0\rangle 
+\frac{2g_C g_L}{\delta_C^2} |G1\rangle
-\frac{2 g_L}{\delta_C}|E0\rangle,\nonumber \\ 
|\tilde{G1}\rangle  &=&  
\left(1-\frac{2 g_C^2}{\delta_C^2}\right)|G1\rangle 
+\frac{2g_C g_L}{\delta_C^2} |G0\rangle
+ \frac{2 g_C}{\delta_C}|E0\rangle,\nonumber\\
|\tilde{E0}\rangle  &=&  
\left(1-2\frac{g_L^2 +g_C^2}{\delta_C^2}\right) |E0\rangle 
+ \frac{2 g_L}{\delta_C}|G0\rangle
- \frac{2 g_C}{\delta_C}|G1\rangle,\nonumber
\end{eqnarray}
and note that up to corrections cubic in $g_{L,C}/\delta_C$ they form
an orthonormal basis of the space spanned by $|G0\rangle$,
$|G1\rangle$, and $|E0\rangle$.  
In this new basis, the Hamiltonian
Eq.~(\ref{eq:Htoy}) takes the diagonal form
\begin{equation}
\label{eq:Htoy1}
 \tilde{H}(m_S=-1)  = \left(\begin{array}{c c c}
-\frac{2 g_L^2}{\delta_C} & 0 & 0\\
0 & \delta_C+\frac{2 g_C^2}{\delta_C} & 0 \\
0 & 0 & \frac{\delta_C}{2} - 2\frac{g_L^2+g_C^2}{\delta_C}
\end{array}\right).
\end{equation}
The absence of any effective coupling between $|\tilde{G0}\rangle$ 
and $|\tilde{G1}\rangle$ in the $m_S=-1$ state for
$\delta_L=\delta_C/2+\Delta$
can be traced back to the equal and opposite 
contributions from coupling the excited state $|E0\rangle$ to the two 
states $|G0\rangle$ and $|G1\rangle$ which are symmetrically arranged in
energy around $|E0\rangle$ in the rotating frame.  
In contrast to this result, we find for the $m_S=0$ state that
\begin{equation}
\label{eq:Htoy0}
 \tilde{H}(m_S=0)  = \left(\begin{array}{c c c}
-\frac{2g_L^2}{\delta_C+2\Delta} & g & 0\\
g & \delta_C+\frac{2 g_C^2}{\delta_C-2\Delta} & 0 \\
0 & 0 & \delta
\end{array}\right),
\end{equation}
with a non-zero amplitude for emitting or absorbing a cavity photon, 
\begin{equation}
\label{eq:Appg}
g =\frac{g_L g_C}{\delta_C-2\Delta}-\frac{g_L
  g_C}{\delta_C+2\Delta} 
= \Delta \frac{g_L g_C}{ (\delta_C/2)^2-\Delta^2},
\end{equation} 
and
\begin{equation}
\delta=\frac{\delta_C}{2} + \frac{g_L^2}{\delta/2+\Delta}-\frac{g_C^2}{\delta_C/2-\Delta}.
\end{equation}
Note that for $\Delta = 0$, the destructive interference of the two
terms in Eq.~(\ref{eq:Appg}) leads to a decoupling, $g=0$.

Using our minimal model, we can also discuss the validity of the
effective Hamiltonian derived using the Schrieffer-Wolff transformation.
The realization of a quantum gate (CPHASE) operation leads to a
time-dependent problem, because the control lasers need to be switched 
on and off to perform the quantum gate. The Schrieffer-Wolff
transformation and use of the obtained effective Hamiltonian for this
time-dependent problem are appropriate if the following two conditions 
are satisfied: (i) Weak coupling (also mentioned above in the text),
more specifically, $g_{L,C}, g\ll\delta_C$,
(ii) adiabatic switching on and off of the laser fields (sufficiently
long ramp time
$\tau_L$) compared to the separation of ground and excited states
(for $m_S=-1$ in the rotating frame), $\tau_{L} \gtrsim \hbar/ \Delta$.

\section{Perturbation analysis}
\label{sec:perturbation-analysis}

In this section we will give an alternative derivation of equation (\ref{Two_NVs_Ground_state_energy}), using conventional time independent perturbation theory.  We are interested in the shift of the ground state of $H_0$,
induced by the perturbation $H_{\rm{int}}$. The matrix element of $H_{\rm{int}}$, that causes the transition from the initial state $|i\rangle$ to the final state $|f\rangle$, is 
\begin{equation}
H^{i\mapsto f}_{\rm{int}}=\langle f|H_{\rm{int}}|i\rangle.
\end{equation}
Thus, the matrix elements of the perturbation are:
\begin{eqnarray}
H^{GG0\mapsto EG0}_{\rm{int}}=H^{GE0\mapsto EE0}_{\rm{int}}=g_{L1}, \\
H^{GG0\mapsto GE0}_{\rm{int}}=H^{EG0\mapsto EE0}_{\rm{int}}=g_{L2} ,
\end{eqnarray}
that account for the interaction between the NV centers and the laser, and 
\begin{eqnarray}
H^{EG0\mapsto GG1}_{\rm{int}}=g_{C1}, \\
H^{GE0\mapsto GG1}_{\rm{int}}=g_{C2}, 
\end{eqnarray}
that account for the interaction between the NV centers and the
cavity. There are also six inverse transitions with the conjugate
matrix elements. 
We consider only the first five energy levels of $H_0$, as we use 
fourth-order perturbation theory and higher energy levels are not excited under this approximation.

One can think of the perturbation to the particular eigenenergy level of $H_{0}$ as arising from transitions that start and end at this level. 
First order processes are thus absent as we have no diagonal terms in
the perturbation. The second order processes are 
\begin{eqnarray}
|GG0\rangle\longmapsto|EG0\rangle\longmapsto|GG0\rangle ,\\
|GG0\rangle\longmapsto|GE0\rangle\longmapsto|GG0\rangle ,
\end{eqnarray}
and the second order energy correction will be
\begin{eqnarray}
\delta E_2&=&-\frac{|g_{L1}|^2}{\delta_{L1}+\Delta ms_1}-\frac{|g_{L2}|^2}{\delta_{L2}+\Delta ms_2}.
\end{eqnarray}
There are no third order processes that would start and end in the
ground state, and therefore, the third order correction to the energy is zero.
Now we include all of the fourth order processes, described by the formula  
\begin{eqnarray}
\delta E_4&=&-\sum\limits_{i,j,k\neq GG0}\frac{H^{GG0\mapsto i}_{\rm{int}}H^{i\mapsto j}_{\rm{int}}H^{j\mapsto k}_{\rm{int}}H^{k\mapsto GG0}_{\rm{int}}}{\langle i|H_0|i \rangle
\langle j|H_0|j \rangle\langle k|H_0|k \rangle}- \nonumber \\
&-&\delta E_2\sum\limits_{i\neq GG0}\frac{H^{GG0\mapsto i}_{\rm{int}}H^{i\mapsto GG0}_{\rm{int}}}{\langle i|H_0|i \rangle^2},
\end{eqnarray}
where we have used that the orbital energy of the state $|GG0\rangle$ can be set to zero.
Also, we have omitted all the terms that contain the diagonal perturbation elements, as those are zero for our system. 
The first term in this equation contains all eight fourth-order processes that exist for this system. The second term is responsible for
renormalization of the perturbed wavefunction. After the calculation, we find
\begin{eqnarray}
\delta E_4&=&
         \frac{|g_{L1}|^4}{(\delta_{L1}+\Delta m_{s1})^3}
      +\frac{|g_{L2}|^4}{(\delta_{L2}+\Delta m_{s2})^3} \nonumber \\
& & -\frac{|g_{L1}|^2|g_{c1}|^2}{(\delta_{L1}+\Delta m_{s1})^2\delta c} 
       -\frac{|g_{L2}|^2|g_{c2}|^2}{(\delta_{L2}+\Delta m_{s2})^2\delta_c} \nonumber \\
& & -\frac{2|g_{L1}g_{L2}|g_{C1}g_{C2}\cos{(\phi_1-\phi_2)}}{(\delta_{L1}+\Delta m_{s1})(\delta_{L2}+\Delta m_{s2})\delta_c}.
\end{eqnarray}
It can easily be seen that $\delta B_1m_{s1}+\delta B_2m_{s2}+\delta
E_2+ \delta E_4$ coincides with the result Eq.~(\ref{Two_NVs_Ground_state_energy})
 obtained in Sec.~\ref{sec:two NVs in a cavity}.

\section{Makhlin invariants}
\label{sec:invariants}

We are interested in producing a two-qubit gate (e.g.  $U_{\rm CPHASE}$)
only up to single-qubit operations, i.e.,
\begin{equation}
U(t) = \exp(-2\pi i t H_{2q}) = (W_1\otimes W_2 ) U_{\rm CPHASE} (V_1\otimes V_2 ),
\end{equation}
with $V_i$ and $W_i$ arbitrary single-qubit unitaries.  
To test whether $U(t)$ 
and $U_{\rm CPHASE}$ are equivalent in this sense, one can use 
two invariants $(G_1, G_2)$ \cite{Makhlin2002} of a 
two-qubit unitary $U$,  defined as
\begin{eqnarray}
G_1 &=& (\textrm{tr} \, m)^2/16 \det U,\\
G_2 &=& ((\textrm{tr}\, m)^2 - \textrm{tr}\, (m^2))/4 \det U ,
\end{eqnarray}
where $m=U_B^T U_B$ and $U_B = Q^\dagger U Q$, with
$Q$ the transformation into the Bell basis,
\begin{equation}
Q = \frac{1}{\sqrt{2}}\left(\begin{array}{c c c c}
1 & 0 & 0 & i \\
0 & i & 1 & 0  \\
0 & i & -1 & 0 \\
1 & 0 & 0 & -i
\end{array}\right).\label{eq:BellQ}
\end{equation}
For the identity operation $U(0) = \openone$, we find
$G_1=1$, $G_2=3$, whereas the CPHASE gate lies in the same
class as the CNOT gate, with $G_1 = 0$, $G_2=1$.
Finding the latter values for $G_1$ and $G_2$ with $U(t)$ for some
time $t>0$ therefore proves that we have generated the CPHASE gate (and with 
this also CNOT gate) up to single-qubit operations.


\end{document}